\documentclass[10pt,a4paper]{article}
\usepackage{graphicx}
\usepackage{amsmath}
\usepackage{url}
\usepackage{bm}
\usepackage{longtable}
\usepackage{adjustbox}
\usepackage{multirow}
\usepackage[subrefformat=parens]{subcaption}
\usepackage{hyperref}
\usepackage{authblk}

\usepackage{array}

\newcolumntype{+}{!{\vrule width 2pt}}

\newlength\savedwidth

\newcommand\thickhline{\noalign{\global\savedwidth\arrayrulewidth\global\arrayrulewidth 2pt}%
\hline
\noalign{\global\arrayrulewidth\savedwidth}}

\newcommand\figref[1]{Fig.~\ref{#1}}
\newcommand\tabref[1]{Table~\ref{#1}}

\begin{document}

\renewcommand{\thefootnote}{\fnsymbol{footnote}}  % temporarily

\title{Money flow network among firms' accounts\\ in a regional bank of Japan}

\renewcommand\Authands{, }
\renewcommand\Affilfont{\small}

\author[1,2]{Yoshi Fujiwara$^\dagger$}
\author[1]{Hiroyasu Inoue}
\author[2]{Takayuki Yamaguchi}
\author[3,4]{\authorcr Hideaki Aoyama}
\author[2,5]{Takuma Tanaka}

\affil[1]{Graduate School of Simulation Studies, University of Hyogo, Kobe 650-0047, Japan}
\affil[2]{Center for Data Science Education and Research, Shiga University, \authorcr
  Hikone 522-8522, Japan}
\affil[3]{RIKEN iTHEMS, Wako, Saitama 351-0198, Japan}
\affil[4]{Research Institute of Economy, Trade and Industry, Tokyo 100-0013, Japan}
\affil[5]{Graduate School of Data Science, Shiga University, Hikone 522-8522, Japan}

\footnotetext[2]{Corresponding author: \texttt{yoshi.fujiwara@gmail.com}}

\date{July 29, 2020}

\maketitle

\begin{abstract} % abstract

In this study, we investigate the flow of money among bank accounts possessed by firms in a
region by employing an exhaustive list of all the bank transfers in a
regional bank in Japan, to clarify how the network of money
flow is related to the economic activities of the firms. The network
statistics and structures are examined and shown to be similar to
those of a nationwide production network. Specifically, the bowtie analysis indicates what we r
efer to as a ``walnut'' structure with
core and upstream/downstream components. To quantify the location
of an individual account in the network, we used the Hodge
decomposition method and found that the Hodge potential of the account has
a significant correlation to its position in the bowtie structure as well as
to its net flow of incoming and outgoing money and links,
namely the net demand/supply of individual accounts.
In addition, we used non-negative matrix factorization to
identify important factors underlying the entire flow of money; it
can be interpreted that these factors are associated with regional
economic activities.
One factor has a feature whereby the remittance source is localized to
the largest city in the region, while the destination is scattered.
The other factors correspond to the economic activities
specific to different local places.
This study serves as a basis for further investigation on
the relationship between money flow and economic activities of firms.

\end{abstract}

\bigskip
\begin{flushleft}
  RIKEN-iTHEMS-Report-20
\end{flushleft}

\renewcommand{\thefootnote}{\arabic{footnote}}
\setcounter{footnote}{0} 

\clearpage
\section*{Introduction}

Determining how money flows among economic entities is an important aspect of 
understanding the underlying economic activities. For example, the
so-called flow of funds accounts record the financial transactions and
the resulting credits and liabilities among households, firms, banks,
and the government (see, e.g., \cite{fofjp}). Another example is
the input-output table, which describes the purchase and sale
relationships among producers and consumers within an economy and
clarifies the flows of final and intermediate goods and services with
respect to industrial sectors and product outputs
(e.g., \cite{iotoecd}). These data are used in 
macroscopic studies, such as those of industrial sectors and aggregated economic
entities.

Recent years have witnessed the increasing emergence of microscopic data. For example, one can
study a nationwide production network, i.e., how individual firms 
transfer money among one another as suppliers and customers for 
transactions of goods and services (see \cite{thebook_mep} and
the references therein). In contrast to the macroscopic studies mentioned
above, microscopic studies can uncover the heterogeneous structure of the
network and its role in economic activities, how the activities are
subject to shocks due to natural disasters \cite{inoue2019flp} and
pandemics \cite{inoue2020pei}, and so forth. However, microscopic data are not exhaustive; although they may cover most active firms, not all the suppliers and customers are recorded. Such records are based on a survey in which a
firm nominates a selected number of important customers and
suppliers. In addition, the transaction amounts are often
lacking; hence, the network is directed but only binary. More importantly,
microscopic and macroscopic data are compiled
and updated annually or quarterly at most (see
\cite{thebook_mep,fujiwara2010large} and the references therein).

To uncover how economic entities such as firms perform economic activities in a real economy, we should ideally study how money flows among firms by using real-time data of bank transfers with exhaustive lists of accounts and transfers. To the
best of our knowledge, such a study has not been conducted thus far, simply because
such data are not available for academic purposes. The present study
precisely performs such an analysis of a Japanese bank's dataset.
The bank is a regional bank, which has a high market share with
respect to the loans and deposits in a prefecture, particularly
supporting financial transactions among the manufacturing firms
located there (according to a disclosure issued by the bank).

The objective of this study is to investigate economic activities via bank
transfers among firms' accounts by selecting all the transfers related
to the firms to uncover how money flows behind the economic
activities. More specifically, we examine the 
network and flow structures, especially the so-called bowtie structure, to
locate the position of individual accounts upstream and downstream of
the entire flow. We quantify the location using the method of Hodge
decomposition of the flow. Furthermore, we find significant factors
underlying the entire flow and interpret them using geographical
information associated with the firms' accounts.

\section*{Data}\label{sec:intro}

Our dataset comprises all the bank transfers that are sent from or
received by the bank accounts in a regional bank. The regional bank is
the largest bank in a prefecture in Japan (mid-sized in terms of its
population (more than a million) and economic activity).
Hereafter, we refer to it as Bank A for anonymity.
The period covered in our study is from March 1, 2017, to
July 31, 2019, i.e., a period of 29 months or 883 days.

During this period, there were 23 million transfers among 1.7
million bank accounts involving a total of 17.4 trillion yen (roughly
160 billion USD or 140 billion Euros). Let us denote a transfer from account
$i$ to account $j$ by $i\rightarrow j$. To focus only on the
firms' accounts in Bank A, we filtered the data such that (i) both $i$
and $j$ are the accounts of Bank A, (ii) both $i$ and $j$
are owned by firms excluding households, and (iii) self-loops
$i\rightarrow i$ are deleted. Point (ii) is important for our
purpose, because our concern here is how money flows and circulates
among firms' accounts, which is considered to be closely related to
the firms' economic activities. The resulting data are summarized in
\tabref{tab:data} (see the rightmost column).

\begin{table}[!ht]
  \centering
  \caption{\textbf{Bank accounts and transfers: summary}}
  \begin{tabular}{|r+r|r|r|}
    \hline
    \multirow{2}{*}{\textbf{Number/Amount}}
    & \multirow{2}{*}{\textbf{Entire data}}
    & \multicolumn{2}{c|}{\textbf{Within Bank A}} \\
    \cline{3-4}
    & & \multicolumn{1}{c}{all} & \multicolumn{1}{|c|}{firms} \\
    \thickhline
    \#Accounts & 1.71 M & 642,411 & 30,613 \\
    \hline
    \#Transfers &  23.06 M & 12,847,963 & 2,409,619 \\
    \hline
    \#Links & 3.13 M & 1,470,107 & 280,864 \\
    \hline
    Transfer (Yen) & 17.43 T & 5.26 T & 2.15 T \\
    \hline
  \end{tabular}
  \begin{flushleft}
    For a transfer $i\rightarrow j$, the column ``Entire data''
    includes the cases in which either $i$ or $j$ is not an account
    of Bank A. The column ``Within Bank A'' corresponds to the case in which both 
    $i$ and $j$ are accounts of Bank A. ``firms'' implies that both the source and the target of a link are firm accounts. M  and T denote million and trillion, respectively.
  \end{flushleft}
  \label{tab:data}
\end{table}

\begin{figure}[tbp]
  \centering
  \includegraphics[width=0.8\textwidth]{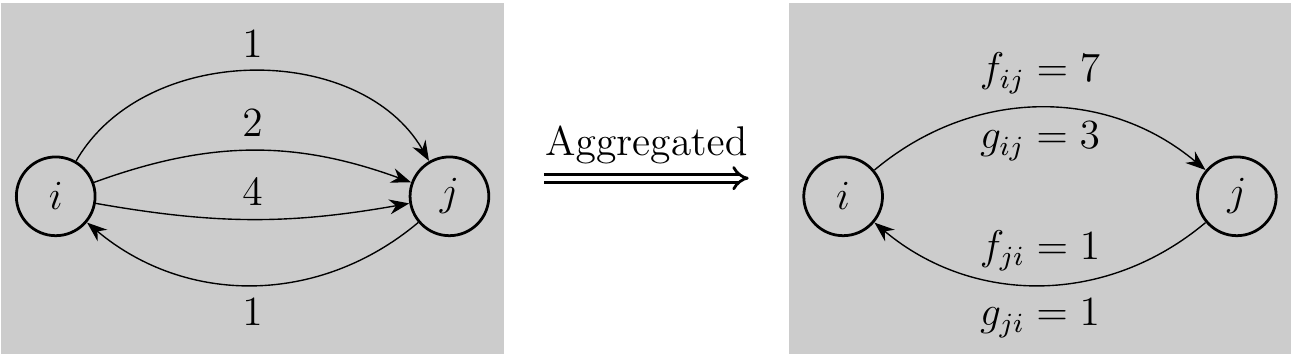}
  \caption{%
    \textbf{Construction of bank-transfer network by aggregation.} %
    How bank transfers are aggregated into links. $i$
    made three transfers (1, 2, and 4) in an arbitrary unit of money to
    $j$, while $j$ made one transfer (1) to $i$ during a certain period. Flow $f_{ij}$ is defined by the total flow of transfers along
    $i\rightarrow j$. Frequency $g_{ij}$ is the frequency of these
    transfers.}
  \label{fig:aggreg}
\end{figure}

Note that multiple transfers $i\rightarrow j$ can exist
for a given pair of $i$ and $j$, because of frequent transfers.
One can quantify the strength of the directional relationship between a
pair of accounts either by the flow of transfers or by their frequency. To do so, we aggregate multiple transfers, if present, into a
single link $i\rightarrow j$ with two types of weights, namely flow
$f_{ij}$ and frequency $g_{ij}$ (see the illustration in
\figref{fig:aggreg}). Hereafter, we use the term \textit{link} for
aggregated transfers.

The number of accounts or nodes in the network is $N=30,613$, while
the number of links is $M=280,864$ after the aggregation (see
\tabref{tab:data}).

The summary statistics of the links' flows $f_{ij}$ and frequencies $g_{ij}$
for all the pairs of accounts $i$ and $j$ are presented in \tabref{tab:stats_link}.
One can observe that the distributions for flow and frequency have large
skewness, implying that a considerable fraction of the money flow is due to
a large amount transferred by a small number of flows.

\begin{table}[!ht]
  \centering
  \caption{\textbf{Summary statistics for links' flows and frequencies}}
  \begin{tabular}{|r+r|r|}
    \hline
    \textbf{Stats.} & \textbf{Flow (Yen)} & \textbf{Frequency} \\
    \thickhline
    Min. & 1 & 1 \\ \hline
    Max. & $3.00\times10^{10}$ & 2,616 \\ \hline
    Median & $0.20\times10^6$ & 3 \\ \hline
    Avg. & $7.65\times10^6$ & 8.58 \\ \hline
    Std. & $1.53\times10^8$ & 19.92 \\ \hline
    Skewness & $92.5$ & $37.8$ \\ \hline
    Kurtosis & $1.25\times10^4$ & $3.49\times10^3$ \\ \hline
  \end{tabular}
  \begin{flushleft}
    Summary statistics of the links' flows and frequencies for all the
    pairs of accounts, where links are aggregated transfers as defined
    in the main text and \figref{fig:aggreg}.
  \end{flushleft}
  \label{tab:stats_link}
\end{table}

\section*{Results and Discussion}

\subsection*{Network of firms' accounts and links of transfers}

\begin{figure}[tbp]
  \centering
  \includegraphics[width=0.85\textwidth]{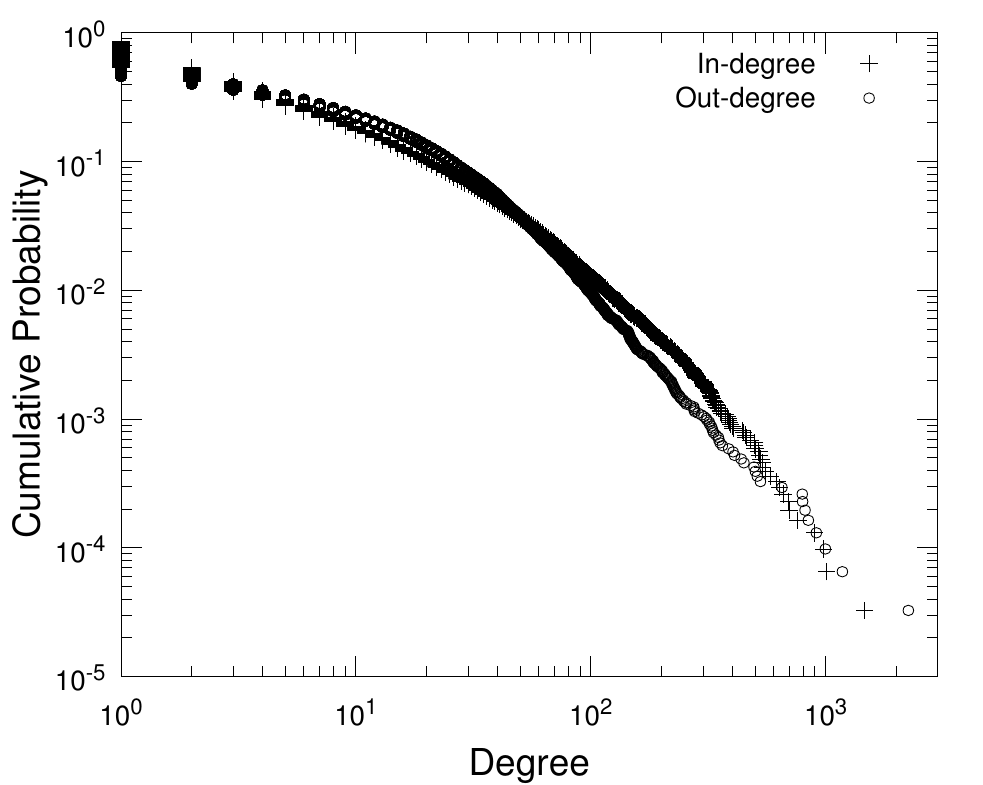}
  \caption{%
    \textbf{Degree distributions for the bank transfer network.} %
    Complementary cumulative distributions for in-degree and
    out-degree, which refer to the number of incoming and outgoing links, respectively, of 
    each account.}
  \label{fig:cpdf_deg}
\end{figure}

First, let us summarize the network structure comprising firms'
accounts as nodes and aggregated transfers as links. We remark that
transfers are aggregated into links as shown in
\figref{fig:aggreg}. The degree is the number of transfers received by or
sent from an account. The number of incoming and outgoing links of an account is
called the in-degree and 
out-degree, respectively. \figref{fig:cpdf_deg} shows the distributions of the in-degree and out-degree as complementary cumulative
distributions. By noting that the total number of accounts is $N=30,613$, we
can see that a small fraction of accounts has a considerable degree, i.e., a
thousand or more links, while most accounts have a limited number
of links. Such hubs are presumably entities 
associated with the local government or the public sector in the
region.

\begin{figure}[tbp]
  \centering
  \includegraphics[width=0.85\textwidth]{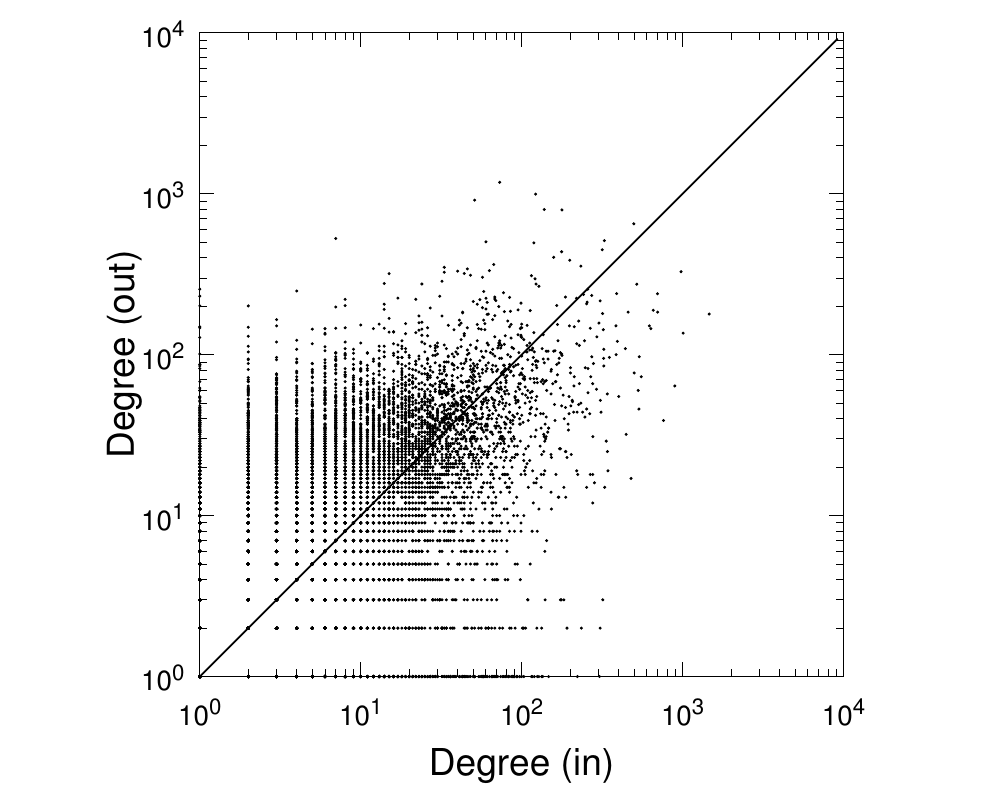}
  \caption{%
    \textbf{Scatter plot for in-degree and out-degree of each account.} %
    Each account as a node, represented as a point, has incoming links
    and outgoing links, the numbers of which are represented by the horizontal
    and vertical axes, respectively. The diagonal line represents 
    the locations where the in-degree and out-degree are equal.}
  \label{fig:deg_corr}
\end{figure}

Because each node has an in-degree and out-degree, we can examine how
they are correlated. \figref{fig:deg_corr} shows the scatter plot for the
in-degree and out-degree of each account. We can observe a
tendency for a positive correlation between the degrees
(Pearson's $r=0.303$ ($p<10^{-6}$);
Kendall's $\tau=0.164$  ($p<10^{-6}$)).
We also observe that there are accounts that have many more incoming
links than outgoing ones (and vice versa), which can be respectively considered as
``sinks'' and ``sources'' with respect to the money flow.

\begin{figure}[tbp]
  \centering
  \includegraphics[width=0.85\textwidth]{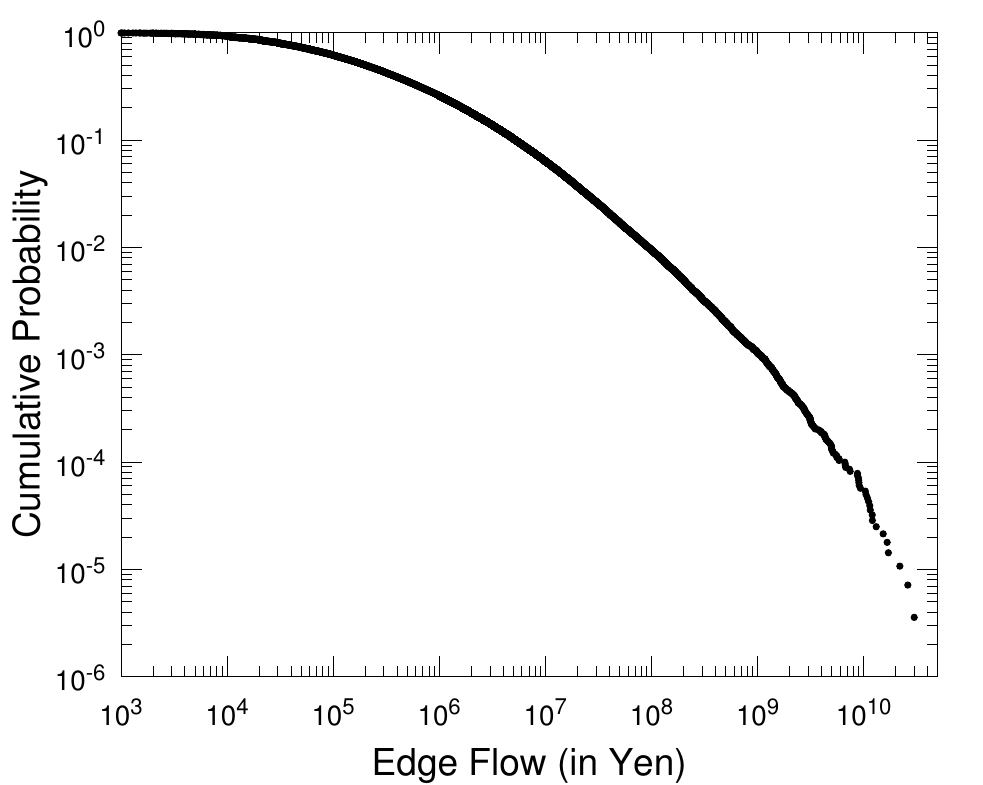}
  \caption{%
    \textbf{Distribution for the flows of links.} %
    Complementary cumulative distributions for the amount of money
    defined by $f_{ij}$ between each pair of accounts $i$ and $j$
    (see \figref{fig:aggreg}).}
  \label{fig:cpdf_e_flow}
\end{figure}

We can observe each link's weights, flow $f_{ij}$, and frequency
$g_{ij}$ (see \figref{fig:aggreg}). \figref{fig:cpdf_e_flow} shows the
complementary cumulative distribution for the flow along each
link. The distribution is highly skewed; there exist a small number of
links that have a large amount of flow exceeding a billion yen---likely important channels with large flows of money. Quantitatively,
0.1\% of the links have flows larger than a billion yen.

\begin{figure}[tbp]
  \centering
  \includegraphics[width=0.85\textwidth]{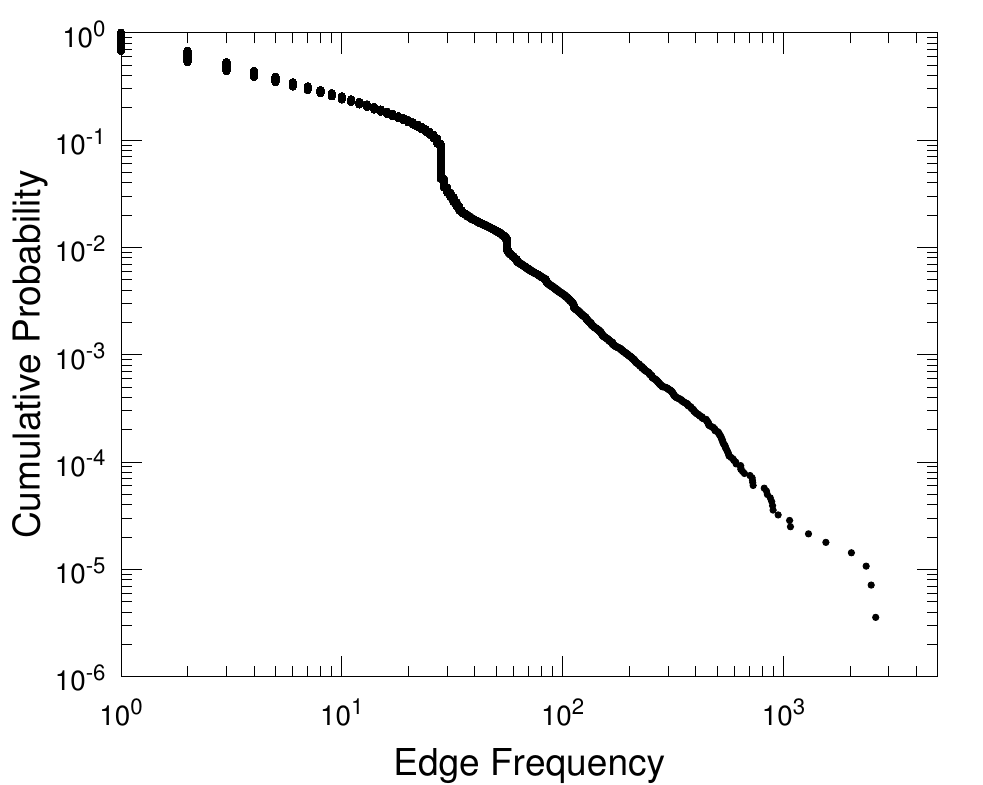}
  \caption{%
    \textbf{Distribution for the frequencies of transfers.} %
    Complementary cumulative distributions for the frequency
    defined by $g_{ij}$ between each pair of accounts $i$ and $j$
    (see \figref{fig:aggreg}).
    We can observe that there are frequency steps around 30 and 60, which are presumed as periodic
    transfers performed once or twice in each month (recall that
    the entire period includes 29 months).}
  \label{fig:cpdf_e_freq}
\end{figure}

\figref{fig:cpdf_e_freq} shows the complementary cumulative
distribution for the frequency along each link. The steps at 30 and 60
on the horizontal axis are considered to correspond to transfers
performed once or twice in each month (recall that the entire period includes 29
months). We can see that 0.1\% of the links have frequencies of 500
or more corresponding to daily transfers on weekdays.

\subsection*{Community analysis}

Communities or clusters in a network are tightly knit groups with high
intra-group density and low inter-group
connectivity \cite{Barabasi16}. Community analysis is useful for understanding how a network
has such heterogeneous structures. We adopt the widely used 
Infomap method \cite{rosvall2008maps,rosvall2011multilevel} to detect
communities in our data.

The results are presented in \tabref{tab:comm}. ``Level'' indicates the level of
communities in a hierarchical tree of communities that are detected
recursively (see \cite{rosvall2011multilevel}). The number of
communities indicates how many communities are detected at the corresponding
level. The label ``irr. comm.'' denotes irreducible communities that
cannot be decomposed further to the next level of smaller communities in
the hierarchical decomposition. For example, 143 of 164 communities at
the first level are irreducible ones, whereas the rest of them are decomposed into
2,327 smaller communities at the next level, and so forth.

\begin{table}[!ht]
  \centering
  \caption{\textbf{Numbers of communities, irreducible communities, and accounts at each level
      of community analysis using Infomap}}
  \begin{tabular}{|r+r|r|r|r|}
    \hline
    \textbf{Level} & \textbf{\#comm.} & \textbf{\#irr. comm.} & \textbf{\#accounts} & \textbf{Ration(\%)}\\
    \thickhline
    1 & 164 & 143 & 355 & 0.012 \\ \hline
    2 & 2,327 & 2,264 & 28,948 & 94.5 \\ \hline
    3 & 215 & 215 & 1,310 & 0.043 \\ \hline
    Total & --- & 2,621 & 30,613 & 100.0 \\ \hline
  \end{tabular}
  \begin{flushleft}
    Each level corresponds to the hierarchical level in the Infomap
    community analysis \cite{rosvall2011multilevel}. A community at a
    level can be decomposed at the next lower level (from top to
    bottom). If a community cannot be decomposed further, it is
    called an irreducible community. The numbers of irreducible
    communities are listed in the third column. The fourth column
    lists the numbers of accounts belonging to these irreducible
    communities at each level.
  \end{flushleft}
  \label{tab:comm}
\end{table}

We find that most of the communities are at the second level because of the number of accounts, and that
most of the accounts (94.5\%) belong the second-level communities.
In our previous study \cite{chakraborty2018hierarchical} on the
application of hierarchical community analysis using Infomap to a
large-scale production network, we showed that a relatively shallow
hierarchy can be observed at the fifth level as the lowest level; in particular,
most firms are included at the second level, exactly as we find
here. This is not surprising, because our data on bank transfers
among firms' accounts should reflect a regional fraction of the entire
production network on a nationwide scale. The finding here is
interesting, because this implies a self-similar structure of the
production network.

\begin{figure}[tbp]
  \centering
  \includegraphics[width=0.85\textwidth]{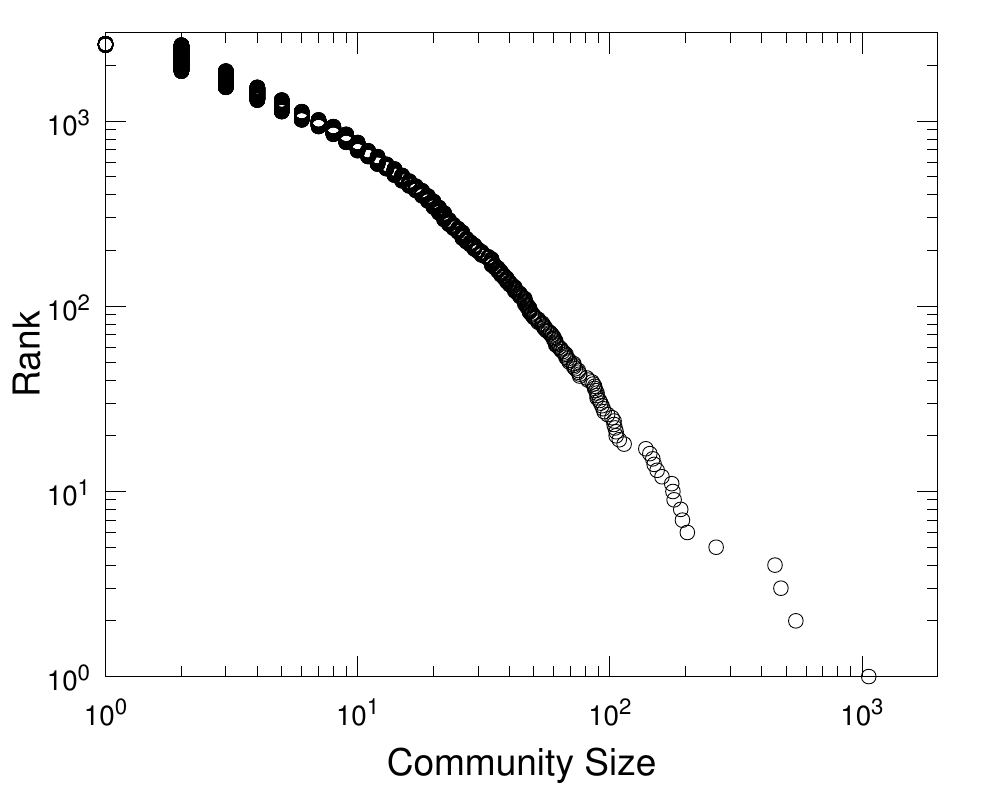}
  \caption{%
    \textbf{Distributions of the sizes of irreducible communities.} %
    Rank-size plot for the sizes of irreducible communities detected
    using the Infomap method at all the levels, where the ranks are in 
    descending order of the size with the lowest rank equal to the
    total number of irreducible communities (see \tabref{tab:comm}).
    The  size  of a community is simply the number of nodes included in the
    community.}
  \label{fig:comm_size-rank}
\end{figure}

\figref{fig:comm_size-rank} shows the distribution of the sizes of irreducible
communities at the lowest level that includes all the
accounts. The size of a community is simply the number of nodes included in
the community. The result indicates that the size of the communities is
highly skewed over a few orders of magnitude. We note
that there exist more than 10 communities with sizes exceeding 100, which correspond to important clusters of economic activities that
depend on geographical sub-regions and industrial sectors. %Detailed analysis is beyond the scope of the present paper, but
We shall discuss this issue in our analysis of
non-negative matrix factorization later.

\subsection*{Bowtie structure}

With respect to the flow of money, the accounts can be located in a
classification of the so-called \textit{bowtie} structure, which was 
first adopted in the study of the Internet \cite{broder2000gsw}. Nodes in a
directed network can be classified into a giant weakly connected
component (GSCC), its upstream side as the IN component, its downstream
side as the OUT component, and the rest of the nodes that do not belong to
any of GSCC, IN, and OUT. In general, they can be defined as follows.
\begin{description}
  \setlength{\itemsep}{0pt}
\item[GWCC] Giant weakly connected component: the largest connected
  component when viewed as an undirected graph. At least one
  undirected path exists for an arbitrary pair of nodes in the
  component.
\item[GSCC] Giant strongly connected component: the largest connected
  component when viewed as a directed graph. At least one directed
  path exists for an arbitrary pair of nodes in the component.
\item[IN] Nodes from which the GSCC is reached via directed paths.
\item[OUT] Nodes that are reachable from the GSCC via directed paths.
\item[TE] ``Tendrils'': the rest of GWCC
\end{description}
Therefore, we have the components such that
\begin{equation}
  \label{eq:bowtie}
  \text{GWCC}=\text{GSCC}+\text{IN}+\text{OUT}+\text{TE}
\end{equation}

For our data of the entire network with $N=30,613$ nodes and
$M=280,864$ links, the GWCC component comprises 30,225 (99.0\%) nodes
and 280,598 (99.9\%) links. The components of GSCC, IN, and OUT are
summarized in \tabref{tab:bowtie}. As can be seen, nearly 40\% of the
accounts are inside GSCC.
Further, 15\% of the accounts are in the upstream
portion or IN, whereas 37\% are in the downstream portion or OUT. These figures
are very similar to those observed in the production network
in Japan in a previous study \cite{chakraborty2018hierarchical}.

The set of three components of GSCC, IN, and OUT is usually referred to as a
``bowtie''; however, we find that the entire shape does not look like
a  ``bowtie'' but like a ``walnut'' in the sense that IN and OUT are two
mutually disjoint thin skins enveloping the core of GSCC
rather than two wings elongating from the center of a bowtie.
In fact, by examining the shortest-path lengths from GSCC to
IN or OUT, we can see that the accounts in the IN and OUT components
are just a few steps away from GSCC as shown in \tabref{tab:bowtie_dist}.
This feature is also similar to the production network on a nationwide scale
(see the walnut structure in \cite{chakraborty2018hierarchical});
however, is different from many social and technological networks such as
the Internet, where the maximum distances from GSCC to IN or OUT are
usually very long (see the original paper \cite{broder2000gsw}).

\begin{table}[!ht]
  \centering
  \caption{\textbf{Bowtie or ``walnut'' structure: size of each component.}}
  \begin{tabular}{|c+r|r|}
    \hline
    \textbf{Component} & \textbf{\#accounts} & \textbf{Ratio(\%)}\\
    \thickhline
    GSCC & 11,543 & 38.2\% \\ \hline
    IN & 4,508 & 14.9\% \\ \hline
    OUT & 11,270 & 37.3\% \\ \hline
    TE & 2,904 & 9.6\% \\ \hline
    total & 30,225 & 100\% \\ \hline
  \end{tabular}
  \begin{flushleft}
    ``Ratio'' refers to the ratio of the number of firms to the total number
    of accounts in GWCC.
  \end{flushleft}
  \label{tab:bowtie}
\end{table}

\begin{table}[!ht]
  \centering
  \caption{\textbf{``Walnut'' structure: shortest distance from GSCC to IN/OUT.}}
  \begin{tabular}{|c|r|r+c|r|r|}
    \hline
    \multicolumn{3}{|c+}{\textbf{IN to GSCC}} &
    \multicolumn{3}{c|}{\textbf{OUT from GSCC}} \\ \hline
    \textbf{Distance} & \textbf{\#accounts} & \textbf{Ratio(\%)} &
    \textbf{Distance} & \textbf{\#accounts} & \textbf{Ratio(\%)} \\
    \thickhline
    1 & 4,346 & 96.41\% &
    1 & 11,051 & 98.06\% \\ \hline
    2 & 144 & 3.19\% &
    2 & 208 & 1.85\% \\ \hline
    3 & 8 & 0.18\% &
    3 & 11 & 0.10\% \\ \hline
    4 & 10 & 0.22\% &
    4 & 0 & 0.00\% \\ \hline
    Total & 4,508 & 100\% &
    Total & 11,270 & 100\% \\ \hline
  \end{tabular}
  \begin{flushleft}
    The left half lists the number of accounts in the IN component connected to
    the GSCC accounts with the shortest distances within 4 at most.
    The right half represents the OUT component similarly.
  \end{flushleft}
  \label{tab:bowtie_dist}
\end{table}

\begin{figure}[tbp]
  \centering
  \includegraphics[width=0.50\textwidth]{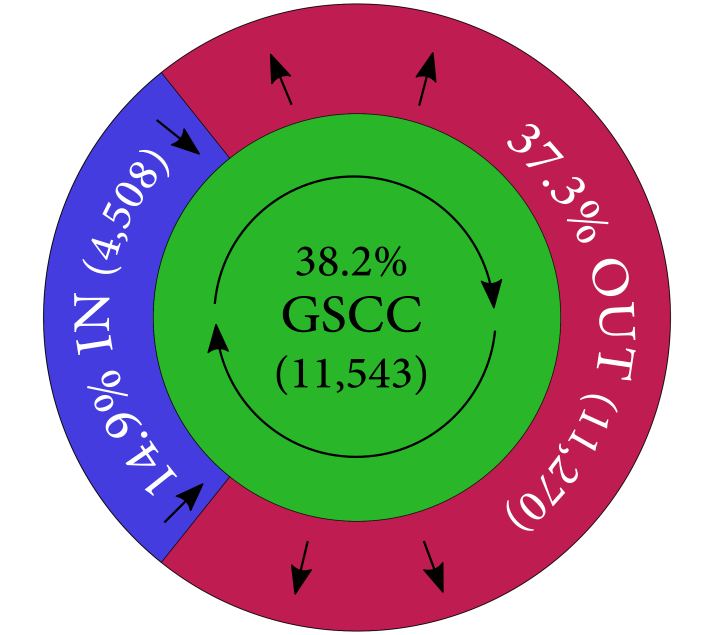}
  \caption{%
    \textbf{Walnut structure: a schematic view.} %
    The so-called bowtie structure reveals that GSCC
    includes nearly
    40\% of all the nodes or accounts, while the IN and OUT
    components include 15\% and 37\%, respectively (see \tabref{tab:bowtie}
    for the details). The prominent features are as follows. (i) The shortest distances
    to IN and OUT from GSCC are quite small, typically
    1 or 2, and 4 at most (\tabref{tab:bowtie_dist}); hence, 
    the ties are not elongated like a ``bowtie'' but rather like
    a ``walnut'' skin. (ii) The nodes in the components of IN and OUT
    are connected to the nodes scattered widely in GSCC.
    See also the study of a supplier-customer network
    \cite{chakraborty2018hierarchical} with similar features.}
  \label{fig:walnut}
\end{figure}

\subsection*{Hodge decomposition: upstream/downstream flow}

Our analysis of the bowtie structure implies that the nodes in IN and
OUT are located in the upstream and downstream sides in the flow of
money. The Hodge decomposition of the flow in a network is a mathematical
method of ranking nodes according to their locations upstream
or downstream of the flow \cite{jiang2011hodge}. This method, also
known as the Helmholtz--Hodge--Kodaira decomposition, has been used to find
such a structure in complex networks (see, e.g., neural networks
\cite{miura2015hodge} and economic networks \cite{kichikawa2018hodge,scirep2020,mackay2020directed}).

First, we recapitulate the method in a manner suitable for our
purpose here. Let $A_{ij}$ denote adjacency matrix of our directed
network of bank transfers, i.e.,
\begin{align}
  \label{eq:def_Aij}
  A_{ij} &=
  \begin{cases}
    1  & \text{if there is a link of transfer from account $i$ to $j$}, \\
    0  & \text{otherwise}.
  \end{cases}
\end{align}
Recall that the numbers of accounts and links are $N$ and $M$,
respectively. We excluded all the self-loops, implying that
$A_{ii}=0$. Each link has a flow, denoted by $B_{ij}$, either of the
total amount of transfers, $f_{ij}$, or the frequency of transfers,
$g_{ij}$ (see \figref{fig:aggreg}), i.e.,
\begin{align}
  \label{eq:def_Bij}
  B_{ij} &=
  \begin{cases}
    f_{ij} \text{ or } g_{ij} & \text{if there is a flow from $i$ to $j$}, \\
    0  & \text{otherwise} .
  \end{cases}
\end{align}
Note that there may be a pair of accounts such that
$A_{ij}=A_{ji}=1$ and $B_{ij}, B_{ji}>0$.
Next, we shall take the frequency of transfers, $g_{ij}$, by
assuming that it represents the strength of the link.

Let us define a ``net flow'' $F_{ij}$ by
\begin{equation}
  \label{eq:def_Fij}
  F_{ij}=B_{ij}-B_{ji}
\end{equation}
and a ``net weight'' $w_{ij}$ by
\begin{equation}
  \label{eq:def_wij}
  w_{ij}=A_{ij}+A_{ji}.
\end{equation}
Note that $w_{ij}$ is symmetric, i.e., $w_{ij}= w_{ji}$, and non-negative, i.e., $w_{ij}\geq 0$ for any pair of $i$ and $j$. We remark
that Eq.~\eqref{eq:def_wij} is simply a convention to consider
the effect of mutual links between $i$ and $j$. One could multiply
Eq.~\eqref{eq:def_wij} by 0.5 or an arbitrary positive number, which
does not change the result significantly for a large network.

Now, the Hodge decomposition is given by
\begin{equation}
  \label{eq:def_hodge}
  F_{ij}=F^{(\text{c})}_{ij}+F^{(\text{g})}_{ij},
\end{equation}
where the \textit{circular flow} $F^{(\text{c})}_{ij}$ satisfies
\begin{equation}
  \label{eq:def_Fcirc}
  \sum_j  F^{(\text{c})}_{ij}=0,
\end{equation}
which implies that the circular flow is divergence-free.
The \textit{gradient flow} $F^{(\text{g})}_{ij}$ can be expressed as
\begin{equation}
  \label{eq:def_Fgrad}
  F^{(\text{g})}_{ij}=w_{ij}(\phi_i - \phi_j),
\end{equation}
i.e., the difference of ``potentials''. In this manner, the weight
$w_{ij}$ serves to make the gradient flow possible only where a link
exists. We refer to the quantity $\phi_i$ as the \textit{Hodge potential}. If
$\phi_i$ is relatively large, the account $i$ is located in the upstream
side of the entire network, while a small $\phi_i$ implies that $i$ is
located in the downstream side of the entire network.

Eqs.~(\ref{eq:def_hodge})--(\ref{eq:def_Fgrad}) can be solved as follows. First, we combine them into the following equation for the Hodge potentials 
$(\phi_1,\cdots,\phi_N) (\equiv\bm{\phi})$:
\begin{equation}
  \label{eq:eqphi}
  \sum_j L_{ij} \phi_j = \sum_j F_{ij} \ ,
\end{equation}
for $i=1,\ldots,N$. Here, $L_{ij}$ is the so-called graph Laplacian and defined by
\begin{equation}
  \label{eq:def_Lij}
  L_{ij}=\delta_{ij} \sum_k w_{ik} - w_{ij} \ ,
\end{equation}
where $\delta_{ij}$ is the Kronecker delta.

It is straightforward to show that the matrix $L=(L_{ij})$ has only
one zero mode (eigenvector with zero eigenvalue), i.e.,
$\bm{\phi}=(1,1,\cdots,1)/\sqrt{N}$. The presence of this zero mode
simply corresponds to the arbitrariness in the origin of $\phi$. We
can show that all the other eigenvalues are positive (see,
e.g., \cite{fujiwara2020hdb}). Therefore, Eq.~\eqref{eq:eqphi} can
be solved for the potentials by fixing the potentials' origin. We
assume that the average value of $\phi$ is zero, i.e.,
$\sum_i \phi_i=0$.

\begin{figure}[tbp]
  \centering
  \includegraphics[width=0.80\textwidth]{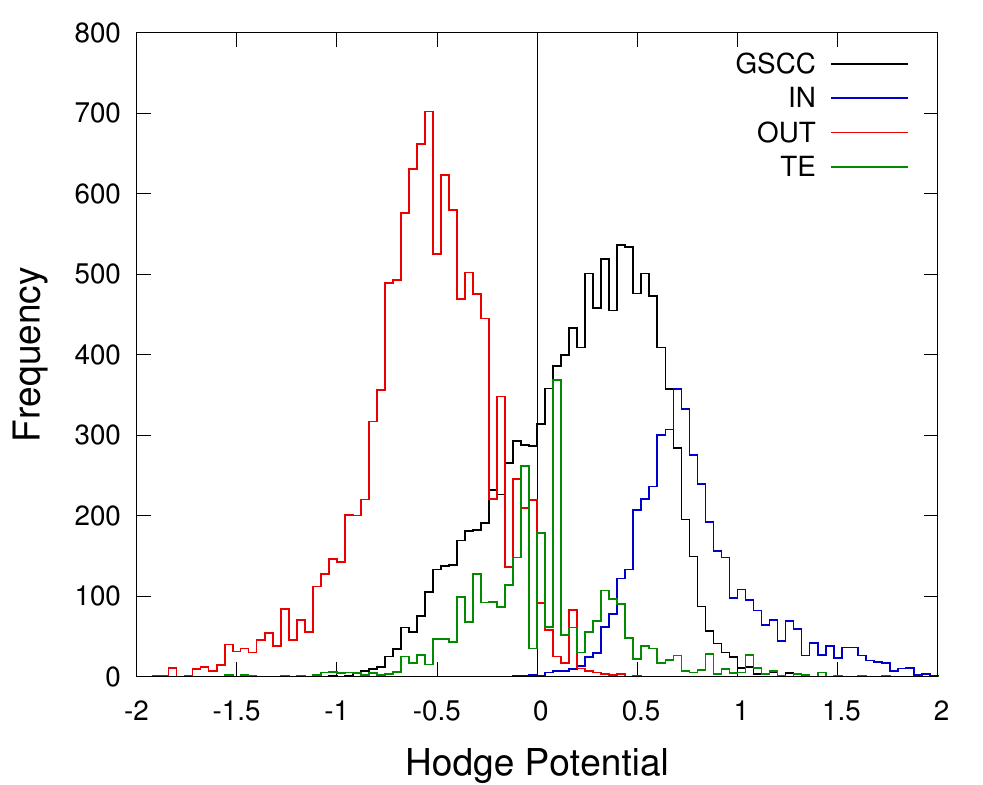}
  \caption{%
    \textbf{Distribution of the Hodge potentials of individual
      accounts.} %
    Distributions as histograms of $\phi_i$ in each
    component of the bowtie or walnut structure \figref{fig:walnut}.
    The horizontal axis represents the value of $\phi_i$ of an individual node or account,
    while the vertical axis represents the frequency in the histogram.
    The black line corresponds to GSCC 
    or the core. The blue and red lines, respectively, correspond to the IN and OUT components
    or upstream and downstream with respect to the core.
    The green line corresponds to TE (tendrils) or the rest of the nodes.}
  \label{fig:pot_binary_hist}
\end{figure}

The Hodge potentials obtained for the entire network of GWCC are shown
in \figref{fig:pot_binary_hist} as the distribution for the potentials
of all the accounts in GWCC (red line). By noting that the average is
zero by definition, we can see that it is a bimodal distribution with
two peaks at positive and negative values, while there are a number of
potential values close to zero (peaks around zero). The nodes in 
TE (tendrils) can be considered to have locations that are not
particularly relevant to upstream or downstream; we can expect that
these nodes mostly have potentials close to zero, as shown by the blue
line, i.e., the result after deleting all the nodes contained in TE's. We
can see that these TE do not contribute to large absolute values of
the Hodge potentials.

\begin{figure}[tbp]
  \centering
  \includegraphics[width=0.85\textwidth]{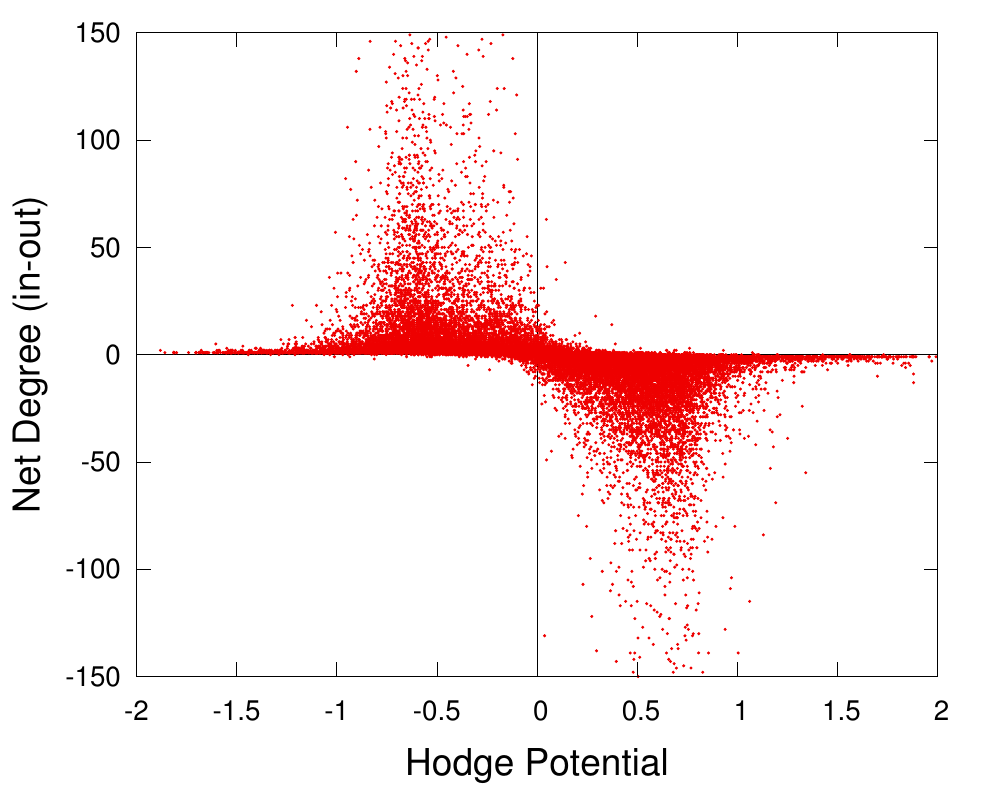}
  \caption{%
    \textbf{Hodge potential and net degree for each node.} %
    Each point represents a node or an account. The net degree is defined
    by the difference between the in-degree and the out-degree of the
    node. If the net degree is positive, the node has more incoming
    links than outgoing ones and vice versa.}
  \label{fig:pot_binary-net_deg}
\end{figure}

\begin{figure}[tbp]
  \centering
  \includegraphics[width=0.85\textwidth]{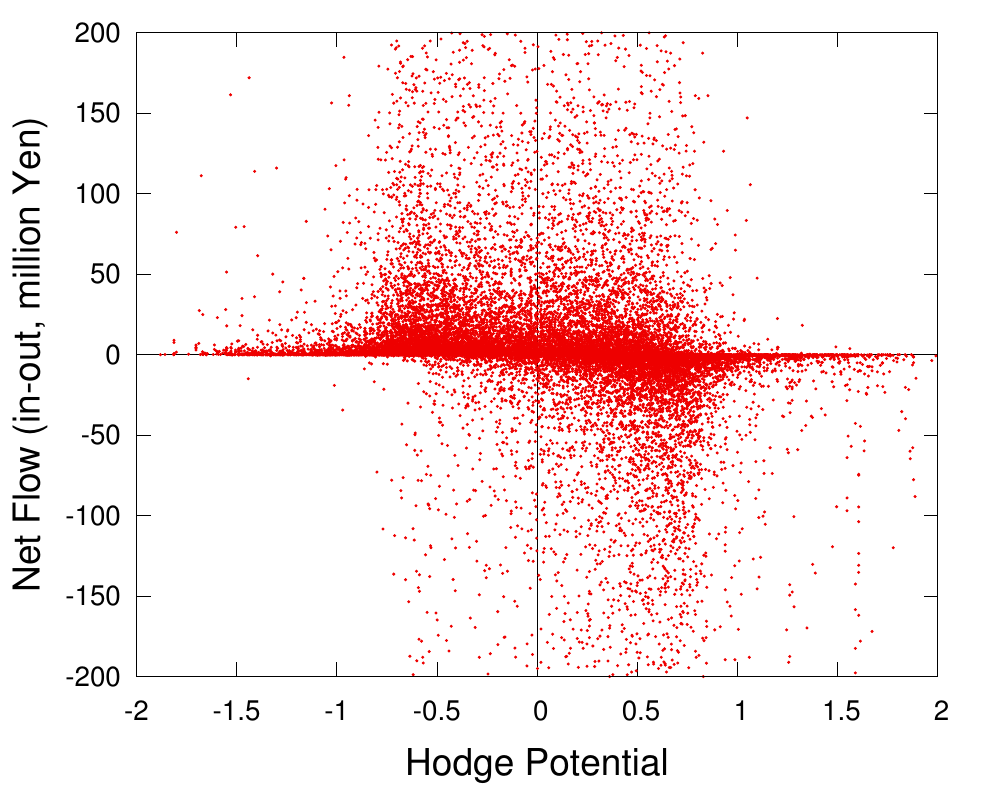}
  \caption{%
    \textbf{Hodge potential and net flow for each node.} %
    This figure is similar to \figref{fig:pot_binary-net_deg} except for the vertical
    axis, which represents the net flow. The net flow is defined by the difference between
    the incoming amount of money and the outgoing one.}
  \label{fig:pot_binary-net_flow}
\end{figure}

It can be expected that there is a correlation between the value of
the Hodge potential and the \textit{net} amount of demand or supply of
money for each node. We can measure the net amount of demand/supply
by examining the in-degree and out-degree of the node, or alternatively,
the in-flow and out-flow of money. \figref{fig:pot_binary-net_deg} and
\figref{fig:pot_binary-net_flow} show the results. We find that if
the potential is positive, the node is located in the upstream
side, and its net degree and flow are negative. If the potential is
negative, the node is located in the downstream side, and its net degree and flow are positive.

This finding can be interpreted as follows. Consider a supplier in the
production network, which supplies its products to a number of 
customers. The supplier has a bank account (or possibly multiple
accounts) that receives money from the customers' accounts as the
supplier's sales. If the supplier is in the upstream side of
the supplier-customer relationship, it is likely that the account is
located in the downstream side of the money flows in this study. As
the supplier not only makes sales but also incurs costs, typically labor
costs, there must be an outgoing flow from its account to be linked with
households and other non-commercial entities, which are not included
in the present study. Consequently, the supplier's account has a
negative net degree and flow, while its Hodge potential is likely
positive. A similar argument would hold for customers in an opposite
way. In other words, our finding is a direct observation of
how the flow of money reflects the economic activities among
the firms' accounts.

\subsection*{Non-negative matrix factorization (NMF): hidden factors of flows}

We would like to show that there are hidden ``factors'' in the entire
flow of the network. By ``factor'', we mean a component that can explain
a significant part of the flow. Alternatively, the entire flow can
be decomposed into only a small number of factors.

In this section, we focus on the geographical information of bank transfers.
Each bank account has an address.
We obtain the latitudes and longitudes of the bank accounts by using geocoding.
Consequently, a bank transfer between two bank accounts has two coordinates of
its remittance source and destination.
We construct a non-negative matrix defined from the frequencies between the geographical areas, and
we adopt NMF to find the hidden factors of geographical structures of the flow.

NMF constructs an approximate factorization of a non-negative matrix
\cite{LeeSeung2000}.
For example, NMF is useful for processing facial images
because it produces parts-based representations of such images \cite{LeeSeung1999}.
To reveal the basic components of the geographical structure of bank transfers,
we apply NMF to a non-negative matrix $V = (V_{mn})$ defined as follows.
We set a square area including the prefecture and
split it into $K \times K$ smaller squares in a lattice pattern, where $K = 100$.
Let $R_{pq}$ be the $(p, q)$ small square area for $1 \le p, q \le K$.
We consider the frequencies of bank transfers between two small square areas.
Let $\alpha (p_1, q_1, p_2, q_2)$ be the frequency of bank transfers from $(p_1, q_1)$ to $(p_2, q_2)$ 
for $1 \le p_1, q_1, p_2, q_2 \le K$,
i.e., using the frequency $g_{ij}$ of transfers from account $i$ to account $j$,
\begin{align}
  \alpha (p_1, q_1, p_2, q_2) = \sum_{\{(i, j) \mid (x_i, y_i) \in R_{p_1 q_1}, (x_j, y_j) \in R_{p_2 q_2} \}} g_{ij},
\end{align}
where $(x_i, y_i)$ is the coordinate of the address of account $i$.
The non-negative matrix $V$ of size $K^2 \times K^2$ is defined by
\begin{align}
  V_{mn} = \log (\max \{1, \alpha (p_1, q_1, p_2, q_2) \}),
\end{align}
where $m = p_1 + (q_1 - 1) K$ and $n = p_2 + (q_2 - 1) K$.
For practical purposes, we convert the frequencies into their logarithmic values to reduce the influence of outstanding values.

NMF gives the approximate factorization
\begin{align}
  V \approx WH \label{eq:NMF}
\end{align}
for some integer $d$,
where $W$ and $H$ are non-negative matrices of size $K \times d$ and $d \times K$, respectively.
We let $d = 10$ from prior knowledge that
the number of local communities in the prefecture is around 10.
Since the $m$th row of $V$ corresponds to bank transfers from $(p, q)$ for $m = p + (q - 1) K$,
the rows of $H$ constitute a basis of bank transfers for the given sources.
Similarly, since the $m$th column corresponds to bank transfers
to $(p, q)$ for $m = p + (q - 1) K$,
the columns of $W$ constitute a basis of bank transfers for the given destinations.
We can regard Eq.~\eqref{eq:NMF} as the approximation of $V$
by the sum of products of these basis vectors.
By letting $w_m$ be the $m$th column vector and $h_m$ be the $m$th row vector, we have
\begin{align}
  V & \approx \sum_{m = 1}^d w_m h_m. \label{eq:NMF-product}
\end{align}
The logarithms of the frequencies of bank transfers in the target area
are decomposed into matrices $w_m h_m$ for $m = 1, \dots, d$.

\begin{figure}[tbp]
  \centering
  \includegraphics[width=0.8\textwidth]{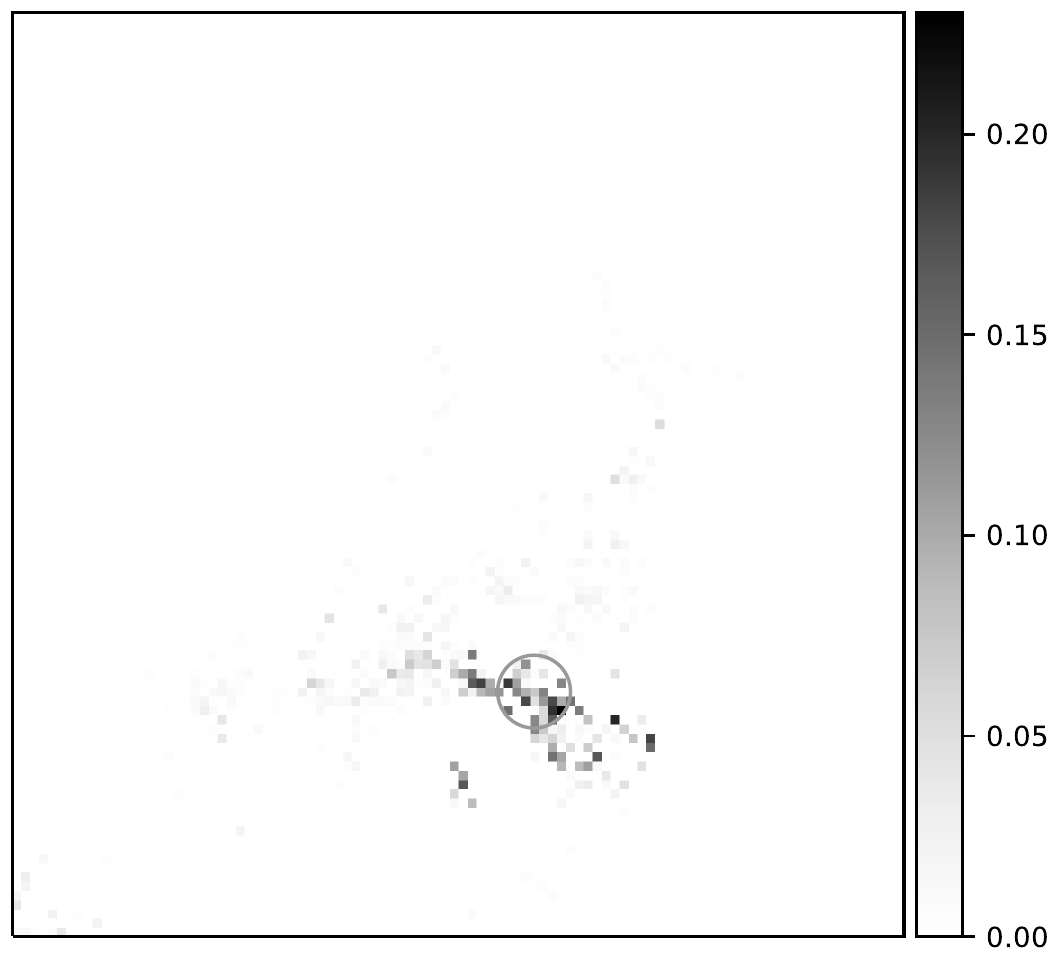}
  \caption{
  \textbf{Normalized basis vector obtained by NMF. The circular area
  has the largest sum of entries of the basis vector included in the circular area.}
  A normalized basis vector such that the sum of entries is one
  is converted into a heatmap whose lattice pattern corresponds to $R_{pq}$.
  The radius of the circular area is 10~km.
  The circular area is $C^{\prime} (v)$ for some basis vector $v$,
  i.e., it is located at a position such that $\beta (\cdot, D(v))$ is maximized.}
  \label{fig:nmf-basis}
\end{figure}

A basis vector $v$, which is a column vector $w_m$ of $W$ or a row vector $h_m$ of $H$,
can be converted to a $K \times K$ matrix $D(v)$, $1 \le p, q \le K$,
on the geographical square area
because an entry of $V$ corresponds to the frequency of bank transfers
between two small square areas.
In other words, $D(v)$ is represented as a heatmap in the geographical area and
\figref{fig:nmf-basis} shows a heatmap of a basis vector.
Since basis vectors seem to indicate geographically localized structures,
to quantify such structures,
we consider a circular area for a basis vector so that the sum of entries of the basis vector
included in the circular area is maximized.
Let $r_{pq}$ be the coordinate of the center of $R_{pq}$ and
let $C_{pq}$ be a circular area whose radius is 10~km and center is $r_{pq}$.
For a $K \times K$ matrix $E = (E_{pq})$ and a circular area $C$, we define
\begin{align}
  \beta(C, E) & = \frac{\sum_{\{ (p, q) \mid r_{pq} \in C \}} E_{pq}}
  {\sum_{ \{ (p, q) \mid 1 \le p, q \le K \} } E_{pq}}.
\end{align}
The proportion $\gamma (v)$ is calculated by
\begin{align}
  C^{\prime} (v) & = \mathop{\rm arg~max}\limits_{ \{C_{pq} \mid 1 \le p, q \le K \}} \beta(C_{pq}, D(v)) \\
  \gamma (v) & = \max_{ \{C_{pq} \mid 1 \le p, q \le K \}} \beta(C_{pq}, D(v)).
\end{align}
The proportion $\gamma$ and the circular area $C^{\prime}$ of a basis vector
are shown in \figref{fig:nmf-basis}.

\begin{figure}[tbp]
  \centering
  \includegraphics[width=0.8\textwidth]{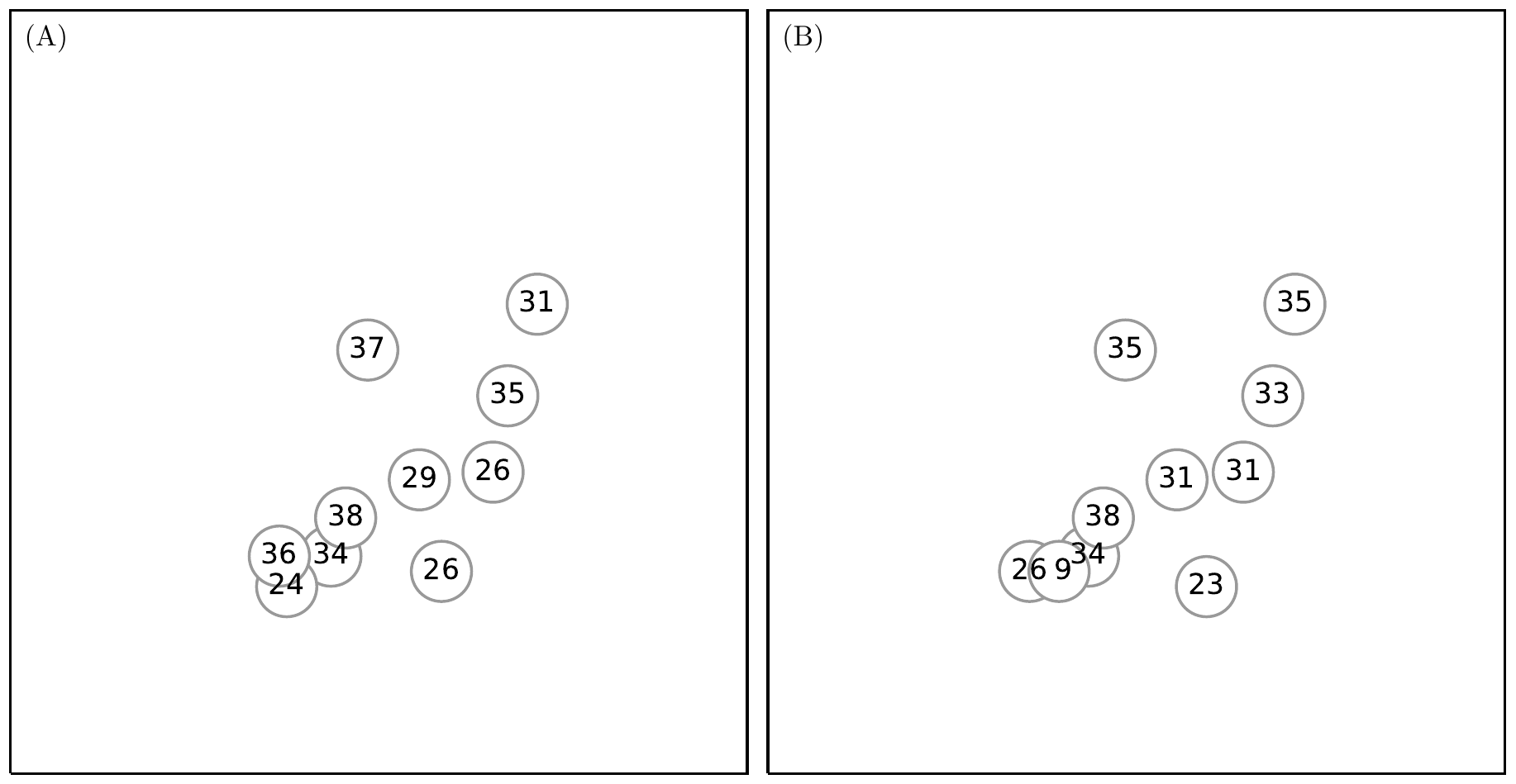}
  \caption{
  \textbf{Circular areas corresponding to the basis vectors
  and proportions of the vector entries included in the circular areas.}
  (A) is drawn from $w_m$, i.e., the basis vectors for sources,
  and the proportions $\gamma (w_m)$,
  while (B) is drawn from $h_m$, i.e., the basis vectors for destinations,
  and the proportions $\gamma (h_m)$ for $m = 1, \dots, d$.}
  \label{fig:nmf-localization}
\end{figure}

\begin{figure}[tbp]
  \centering
  \includegraphics[width=0.8\textwidth]{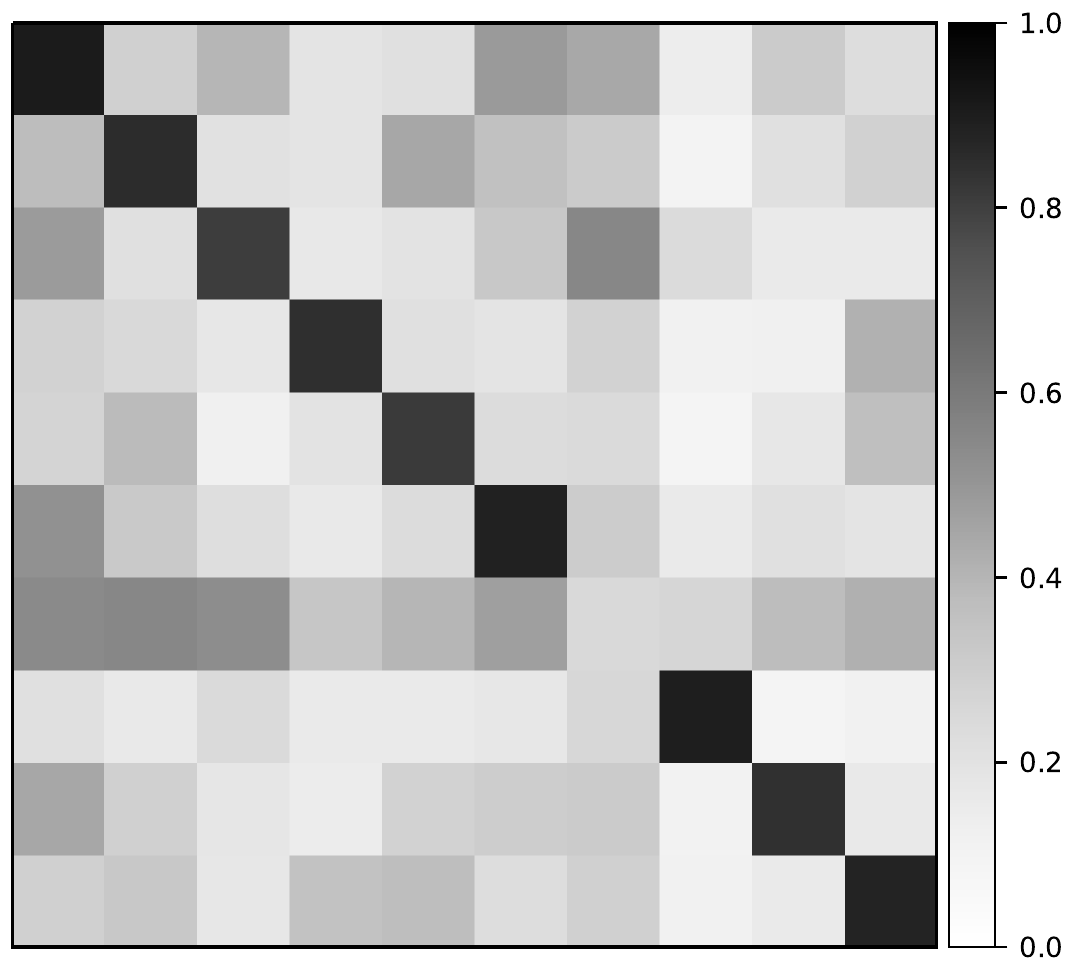}
  \caption{
  \textbf{Cosine similarities between basis vectors.}
  The vertical axis represents the indices of $h_s$, i.e., the $s$th row vector of $H$,
  and the horizontal axis represents the indices of $w_t$, i.e., the $t$th column vector of $W$.
  The index of the top left square is $(s, t) = (0, 0)$.}
  \label{fig:nmf-inner-product}
\end{figure}

The panels (A) and (B) in \figref{fig:nmf-localization} show
the proportions $\gamma$ of all the basis vectors of sources and destinations.
The proportions are more than 23\% except for one basis vector in both panels of the source and destination; therefore, most basis vectors of bank transfers are localized geographically.
Since the positions of the circular areas are around the centers of cities,
geographically localized properties are thought to reflect the economic activity in local areas.

\figref{fig:nmf-localization} suggests that the basis vectors of
the source and destination are similar to each other.
To clarify this, \figref{fig:nmf-inner-product} shows a matrix of cosine similarities
between a basis vector of the source and a basis vector of the destination,
where the cosine similarity of $w_m$ and $h_n$ is calculated by
\begin{align}
  \frac{w_m \cdot h_n}{\|w_m\|\|h_n\|},
\end{align}
where $w_m \cdot h_n$ is the inner product of $w_m$ and $h_n$ and
$\|\cdot\|$ is the Euclidean norm of a vector.
All the diagonal entries except for one are 1's, i.e.,
the $m$th basis vector $h_m$ is similar to the $m$th basis vector $w_m$
except for $m = 7$.
These basis vectors correspond to basis vectors having geographically localized properties
in \figref{fig:nmf-localization},
and the similarities of pairs of basis vectors imply that
both incoming and outgoing bank transfers for a local area have similar patterns.

We can also interpret the seventh basis vectors of the source and destination that do not have similarities.
The seventh basis vector of the source is localized to the largest city in the prefecture
and the seventh basis vector of the destination is scattered throughout the prefecture.
This means that the pair of these basis vectors corresponds to bank transfers
from the largest city to the local areas.
Therefore, Eq.~\eqref{eq:NMF-product} for our data gives decompositions that describe
bank transfers in local areas and bank transfers between the largest city and local areas.

Finally, we state the results of NMF with different values of $d$.
To investigate the changes in the basis vectors that occur according to $d$,
we apply NMF to $V$ with $d = 5, \dots, 15$.
In all the cases, most of the basis vectors are geographically localized and
form source and destination pairs that are similar to each other and
correspond to bank transfers in local areas.
All the basis vectors are localized for $d$ less than 7, and
there is a pair of basis vectors corresponding to bank transfers
between the largest city and local areas for $d$ greater than or equal to 7.
For all the values of $d$ that we have examined,
the basis vectors correspond to either bank transfers in local areas or
bank transfers between the largest city and other local areas.

\section*{Conclusion}

We studied an exhaustive list of bank accounts of firms and remittances
from source to destination within a regional bank with a high market
share of loans and deposits in a prefecture of Japan. By studying such a network
of money flow, we could uncover how firms conduct the underlying economic activities
as suppliers and customers from the upstream side to the downstream side of the
money flow. We aggregated the remittances that occurred for each
pair of accounts as a link during the period from March 2017 to
July 2019 (i.e., approximately two and a half years), which comprises 
30K nodes and 0.28M links. We found that the
statistical features of the network are actually similar to those
of a production network on a nationwide scale in Japan
\cite{thebook_mep}, but with greater emphasis on the regional aspects.

The bowtie analysis revealed what we refer to as a ``walnut''
structure in which the core and upstream/downstream components are tightly
connected within the shortest distances, typically at a few
steps. By quantifying the location of the individual account of a firm using the
method of Hodge decomposition, we found that the Hodge potential of
each node can describe the location in the entire flow of money from
the upstream side to the downstream side, well characterized by the values of the
potential. In particular, there is a significant correlation between
the Hodge potentials and the net flows of incoming and outgoing money
and links as well as the potentials and the walnut structure. This
implies that we can characterize the net demand/supply of each
node and decompose the flows into those due to the
difference in potentials as well as divergence-free flows.
Furthermore, by using non-negative matrix
factorization, we uncovered the fact that the entire flow can be
considered as a combination of several significant factors. One
factor has a feature whereby the remittance source is localized to
the largest city in the region, while the destination is scattered. The other factors correspond to the economic activities
specific to different local places, which can be interpreted as 
local activities of the economy.

We can consider several points that remain to be studied separately from the
present work. While we aggregated the entire period in this paper, it
would be interesting to determine how the network changes with time by examining the
time-stamps recorded in every remittance. At time scales of days, 
weeks, and months, it is quite likely that there are intra-day, weekly,
and seasonal patterns of activities. More interestingly, under mild
changes in the booms and busts of the regional economy on a relatively
long time scale, the economic agents might change their behaviors
possibly by changing peers in the transactions. Alternatively, under 
sudden changes due to natural disasters or pandemics, the agents can change
their usual patterns abruptly. In other words, these are important
aspects of a temporally changing network.

In addition, further investigation of the aspect of money
flow amounts is warranted in the sense that the dominant driving force likely comes
from ``giant players'' who demand or supply a large amount of money.
Moreover, it would be interesting to select them in a subgraph by
choosing only links with flow amounts that are larger than
a certain threshold. These topics will be studied in our future work.

\bigskip
% ----------------------------------------
\section*{Acknowledgements}
We would like to thank Bank A for giving us an opportunity to
study such a unique and valuable dataset. %
We are also grateful to Yoshiaki Nakagawa (Center for Data Science
Education and Research, Shiga University) for insightful discussions.

\section*{Funding}
This work was supported in part by MEXT as Exploratory Challenges on
Post-K computer (Studies of Multi-level Spatiotemporal Simulation of
Socioeconomic Phenomena), the project ``Large-scale Simulation and
Analysis of Economic Network for Macro Prudential Policy'' undertaken
at the Research Institute of Economy, Trade and Industry (RIETI), and 
JSPS KAKENHI Grant Numbers 17H02041, 19K22032, and 20H02391.

\section*{Availability of data and materials}
The dataset is available in a collaborative scheme upon request
to TT and YF at Shiga University.

\section*{Competing interests}
The authors declare that they have no competing interests.

\section*{Author's contributions}
All authors contributed equally. All authors read and approved the final manuscript.

% ----------------------------------------
\bibliographystyle{unsrt}
\bibliography{refs}

\end{document}